\documentclass[twocolumn,aps,pra,amsmath,amssymb,showpacs,superscriptaddress]{revtex4-1}

\usepackage{graphicx}
\usepackage{graphics}
\usepackage{dcolumn}
\usepackage{bm}
\usepackage{color}
\usepackage{soul}
\usepackage{textcomp}
\usepackage{soul}
\usepackage{rotating}
\usepackage{setspace}
\usepackage[mathlines]{lineno}
\usepackage{enumerate}

\usepackage{microtype}
\usepackage[breaklinks=true,hyperindex=true,pdftitle={Quantum Driven Dissipative Parametric Oscillator in a Blackbody Radiation Field},
colorlinks=true,pagebackref=false,citecolor=blue,plainpages=false,pdfpagelabels,
linkcolor=blue,urlcolor=blue]{hyperref}

\begin{document}


\title[Incoherent Excitation of Thermally Equilibrated Open Quantum Systems]{
Incoherent Excitation of Thermally Equilibrated Open Quantum Systems
}

\author{Leonardo A. Pach\'on}
\affiliation{Instituto de F\'{\i}sica, Universidad de Antioquia, AA 1226
Medell\'in, Colombia}
\affiliation{Chemical Physics Theory Group, Department of Chemistry and
Center for Quantum Information and Quantum Control,
\\ University of Toronto, Toronto, Canada M5S 3H6}

\author{Paul Brumer}
\affiliation{Chemical Physics Theory Group, Department of Chemistry and
Center for Quantum Information and Quantum Control,
\\ University of Toronto, Toronto, Canada M5S 3H6}

\begin{abstract}
Under natural conditions, excitation of biological
molecules, which display non-unitary open system dynamics, occurs
via incoherent processes such as temperature changes
or irradiation by sunlight or moonlight. The dynamics of such processes is 
explored \emph{analytically} in a non-Markovian generic model. Specifically, a system S in 
equilibrium with a thermal bath TB is subjected to an external incoherent
perturbation BB (such as sunlight) or another thermal bath TB$^\prime$, 
which induces time evolution in (S$+$TB). Particular focus is on (\textit{i}) the extent
to which the resultant dynamics is coherent, and (\textit{ii}) the role of 
``stationary coherences", established in the (S$+$TB) equilibration, in
response to the second incoherent perturbation. Results for systems
with parameters analogous to those
in light harvesting molecules in photosynthesis show that the resultant
dynamical behavior is incoherent beyond a very short response to the turn-on
of the perturbation.
\end{abstract}

\date{\today}


\maketitle

\section{Introduction}
The dynamics of open quantum systems (e.g., molecules of interest in contact with an 
environment) in the presence of an external perturbation is of great current interest and 
applicability. 
Two general circumstances can be envisioned, one where the external perturbation is 
``designed", the second where it occurs naturally. 
Within the framework of electromagnetic perturbations, popular examples of the former
include pulsed laser studies of molecular dynamics in the laboratory \cite{EC&07,PN&06},
laser based scenarios for quantum control and quantum information, various
spectroscopies, etc. 
These artificial light sources are often characterized by pulses of short temporal duration 
and  significant coherence times.

By contrast, natural light induced processes such as vision or photosynthesis are induced 
by virtually stationary chaotic blackbody light sources such as the sun, which have significantly 
different properties \cite{JB91,HB07,HB11,MV10,BS12b,HSB12}. 
Similarly, natural processes, like ion transport through membranes \cite{XC&11} may 
be induced by temperature changes. 
Both perturbations, temperature change and excitation by natural light, are fully incoherent. 
That is, they are both associated, as outlined later below, with a perturbation described 
by a density matrix that is a mixture of stationary states.

{A proper treatment of the systems described above requires that the system S of 
interest be first equilibrated with a background thermal bath TB, after which it is subjected 
to a second perturbation comprising a second thermal bath (denoted TB$^\prime$) 
or blackbody radiation (denoted BB). 
Our focus here, in this two step process, is on the role of coherences in the dynamics.}

One motivation for this work lies in recent studies of
coherent quantum dynamics in model photosynthetic light harvesting
systems \cite{EC&07,CW&10} and in vision \cite{PN&06,HB07,HB11}.
Coherent dynamics of this kind is
observed in experiments in which the system is excited with coherent
laser light, and the timescales for the decay of the coherence, generated
by the system interacting with its environment, is then measured.
Principal among the observations on molecules involved in photosynthesis
is that the coherent dynamics associated
with electronic energy transfer persists, at room temperature,
 on considerably
longer time scales (e.g.,  400 to 2000~fs) than is
expected from earlier results on other systems \cite{HR04,HR04b,FB08,FB11},
and from theories of decoherence \cite{Sch07}.
A number of computational
results have been obtained, and theories advanced, as to why such longevity
occurs. For example, we have identified a number of physical conditions
under which such long lived electronic coherence persists \cite{PB11,PB12}.

The relevance of these results for realistic photosynthetic systems,
or for the operation of devices that mimic photosynthesis, however,
depends heavily
on the relationship between the dynamics observed in the experiments, which use
pulsed coherent light, and dynamics
under natural light, such as that from the sun.
Both recent \cite{BS12b} as well as earlier studies \cite{JB91} on this relationship for \textit {isolated
molecules} (i.e. molecules that are not in contact with an environment)
show that whereas pulsed coherent light induces dynamics in the molecule,
natural stationary chaotic light does not. Rather, irradiating an isolated molecule with
such natural light over long natural time periods
creates a \textit{stationary} mixture of molecular eigenstates.

Natural light absorbing molecules of interest in, e.g., photosynthesis and vision are not, however,
isolated. 
Rather they are in contact with an external environment through which decoherence 
and relaxation occur. 
Hence, there is a need to understand the response of a system$+$bath that is subject to an incoherent
excitation. 
A study of Retinal excitation \cite{HB11} under these conditions confirmed that stationary 
eigenstates resulted in this open system as well. 
An alternate study on photosynthetic systems \cite{MV10} simultaneously in contact 
in contact with both (TB+BB) 
noted the appearance of off-diagonal elements of the subsystem density matrix S 
and regarded them as ``stationary coherences", a concept elucidated further below. 
The relation of these stationary coherences to dynamics was, however, unclear. 
Hence, clarifying the situation regarding open systems under incoherent 
perturbation is, therefore, well motivated.

As noted above and as emphasized, in this work it is important
to recognize that the natural processes noted above occur
in two distinct steps.
That is, in the first step the subsystem S thermally equilibrates with
its surrounding bath, yielding a result that we denote as (S$+$TB). 
In the second step this equilibrated system is placed in contact with a source 
of incoherent light (BB) or another thermal bath TB$^\prime$ with
which the system interacts, giving an (S$+$TB)$+$BB  [or (S$+$TB) $+$TB$^\prime$] system.
As described in detail below in Sec.~\ref{statcoh}, the first generates
``stationary coherences". 
The nature of
the subsequent relaxation dynamics and the role of the stationary
coherences arising from the equilibration of S with TB in the
subsequent dynamics of (S$+$TB)$+$BB [or of (S$+$TB)$+$TB$^\prime$] are described below.

The  existence  of  \textit{substantial} stationary coherences
requires  strong  system-environment coupling (whether
dissipative  or  not) \cite{Blo65,PB&81,*Blo86,*Rot87},  and as
we  will  show,  are  largest  in  the low temperature regime.
This regime  of  strong  coupling  and  low  temperature  is
relevant in the context of,  e.g.,  electronic  energy
transfer \cite{PB11,PB12}.  Being  in  this regime prevents the
application  of  simplifying  approximations  to the evolution
of  the  density  matrix,  such as the Markovian
approximation \cite{SB11}  or  secular  approximation.  That is
why  a  thorough  analysis  of  these  off-diagonal  terms and
their  influence  on  the  dynamics have been elusive. Here we
show, using an analytic model 
that  these  off-diagonal  terms  can  contribute to the
dynamics  of  the  populations only when they enter as initial
correlations  between  the  system  and the thermal bath, and that in
realistic systems these contributions are small.

Note for clarity below, the specific characteristics of the natural process that are of
interest. 
That is, we are focusing on molecular systems that are, as in natural photosynthetic 
processes, irradiated by blackbody radiation  for time 
scales that are far far longer (e.g. hours) 
than the inverse of the molecular energy level spacing, which would define the time scale 
for coherent molecular dynamics.
Results of this study are found to extend the result previously obtained for the isolated
system to the open system. 
That is, ``natural incoherent chaotic light" is shown to be incapable of inducing 
coherent dynamics in either isolated or open systems.

Note that the dynamics examined here takes place on a single electronic surface.
The extension of this analytic influence functional approach to non-adiabatic
processes is in progress. However, the general conclusions obtained here are expected
to hold in the case of excitation from one electronic surface to another.

The paper is organized as follows: Sec.~\ref{statcoh} introduces features
of the first thermalization step, i.e. the process of S$+$TB relaxing, with
a focus on the concept of ``stationary coherences". Section \ref{firstrelax}
provides the Hamiltonian under consideration and associated computational
results for the (S$+$TB) relaxation. 
The case of (S$+$TB) interacting with a second thermal bath is contained in Sec.~\ref{TBPcase}, 
whereas the case where the perturbation is incoherent light is discussed in Sec.~\ref{BBcase}. 
A short discussion and summary is provided in Sec.~\ref{disc}.

\section{Equilibration of (S$+$TB) and ``Stationary Coherences"}
\label{statcoh}
The essence of open quantum systems lies in the interaction between a part of
interest, ``the system", with a component which is not of interest ``the environment"
or ``bath". 
Consider then the two components, a system $\hat{H}_{\mathrm{S}}$ with eigenbasis 
$\{n_i\}$ defined on the Hilbert space $\mathcal{H}_{\mathrm{S}}$ and (in this section) 
the thermal bath described by the Hamiltonian $\hat{H}_{\mathrm{TB}}$ with eigenbasis 
$\{N_i\}$ defined on the Hilbert space $\mathcal{H}_{\mathrm{TB}}$. 
In the absence of an interaction between them, the total
Hilbert space $\mathcal{H} = \mathcal{H}_{\mathrm{S}} \otimes \mathcal{H}_{\mathrm{TB}}$ 
will be diagonal in the basis $\{n_i \otimes N_i\}$. 
However, if the systems interact via a coupling term $\hat{H}_{\mathrm{ST}}$, then $\mathcal{H}$ 
is no longer  diagonal in this basis. 
To obtain information on the system, we would then trace over the degrees of freedom of 
the bath. 
It is then clear that the remaining sector $\mathcal{H}_{\mathrm{S}}$ is no longer diagonal  
in the $\{n_i\}$ basis. 
Rather, the new \emph{effective} eigenbasis should ``know about the thermal bath"; therefore 
the resultant effective system basis should be a function of the coupling to the bath and of the
temperature.

In general, one is not interested in studying the dynamics in terms
of this coupling-and-temperature dependent basis, but rather in describing the 
dynamics in the system eigenbasis of $\hat{H}_{\mathrm{S}}$. 
Viewed in this basis, the coupling to a bath generates off-diagonal elements 
in the system density matrix. 
Once the (system$+$bath) have relaxed to equilibrium these off-diagonal
elements do not change with time, and are not associated with dynamics. 
Rather, they can be termed ``stationary coherences". 
They are just a manifestation of our focus on the system rather than on the 
coupled system$+$bath.  

An alternate perspective on these stationary coherences is that strong coupling
to the environment causes overlap of homogeneous line shapes associated with 
different energy states of system S.

Given that the equilibrated S+TB composite is the natural state in which
one finds the system S in the cases of interest, the relevant questions here are then:
(\textit{i}) what are the nature and times scales of the
dynamics when this equilibrated (system$+$bath) is exposed to an 
incoherent perturbation, and (\textit{i}) what is the role of the stationary 
coherences in this subsequent dynamics? 
For example, it has been suggested \cite{MV10}, but 
not explored quantitatively, that such terms are capable of 
inducing coherent dynamics when the equilibrated S$+$TB is subject to an 
additional perturbation.

Below  we  show how to properly address the incoherent excitation of 
a system S in a thermal environment by, in contrast
with  previous  work  \cite{HB07,HB11,MV10},  considering  the
excitation in two steps.  In
particular,  using  an analytically soluble model valid in the
whole range of parameters, we study the role  of
environment-generated   coherences   during:  (\emph{i})
thermalization  of  an  initially isolated central system S in
contact with a thermal  bath  TB,  and  (\emph{ii})  the
subsequent  dynamics  induced  by  the  presence  of  a second
thermal  bath  TB$^\prime$  at  different  temperature  or  by
blackbody  radiation  BB. In the absence of initial coherences
in  the  system  eigenbasis,  we  show in situation (\emph{i})
that,  although  time dependent off-diagonal elements could be
detected,  they  are unrelated to the dynamics of the diagonal
terms. In  situation  (\emph{ii}),  coherences  having  been
initially  generated  in  S  by TB contribute naturally to the
dynamics  of  the  populations of S toward the new steady state.

\section{Coherences and dynamics toward thermalization}
\label{firstrelax}
\begin{figure}
\includegraphics[width = 0.75\columnwidth]{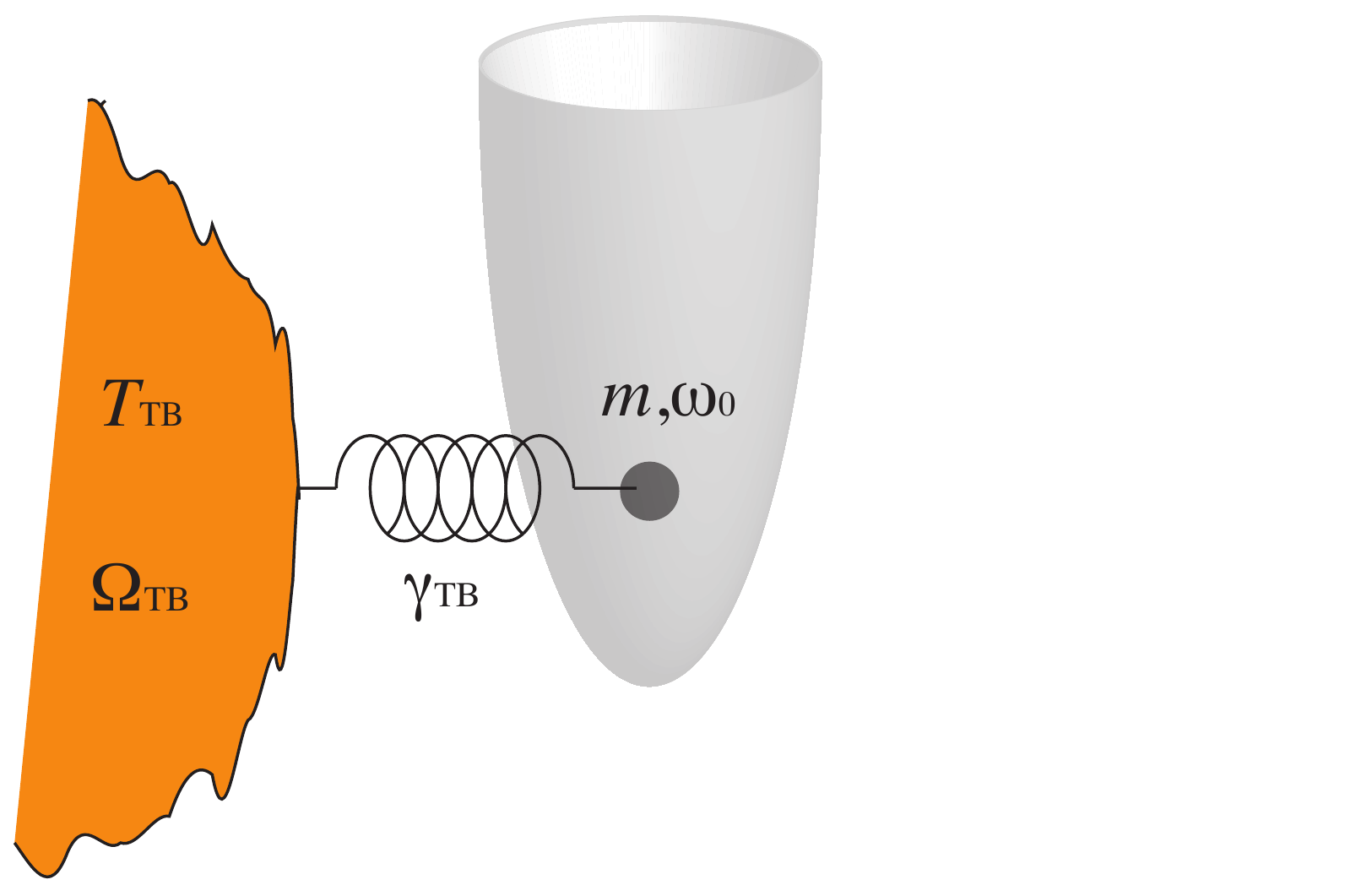}
\caption{The system S, initially described by $\hat{\rho}_{\mathrm{S}}(0)$, is coupled to
the thermal bath TB at temperature $T_{\mathrm{TB}}$ and frequency cutoff $\Omega_{\mathrm{TB}}$.
$\gamma_{\mathrm{TB}}$ denotes the strength of the cooling.}
\label{fig:onebath}
\end{figure}

The process described above is not a particular feature of light-harvesting systems,
but is a generic feature of open quantum systems and is therefore  ubiquitous. 
In order to appreciate this \emph{in detail}, and to extract the relevant features of the underlying 
physical situation, we consider a model: a harmonic oscillator immersed in a dissipative
environment TB \cite{CL83,GSI88,FLO85,*FLO87,*FLO88}. 
Although an idealization, it encompasses a reasonable description of a wide variety of objects 
in nature such as low energy vibrational molecular modes, in addition to artificial ones like 
nanomechanical oscillators, optical and microwave cavities, and movable mirrors \cite{GPZ10}. 
The Hamiltonian of the system$+$environment can be written as
$\hat{H} = \hat{H}_{\mathrm{S}} + \hat{H}_{\mathrm{TB}} + \hat{H}_{\mathrm{ST}}$,
where $\hat{H}_{\mathrm{S}}$ is the Hamiltonian of the unperturbed oscillator, 
$\hat{H}_{\mathrm{TB}}$ is the Hamiltonian of the thermal bath TB, and $\hat{H}_{\mathrm{ST}}$ 
describes the interaction of the system with TB. 
In particular, we choose 
\begin{equation}
 \label{eq1}
\hat{H} =  \hat{H}_{\mathrm{S}} + 
 \sum_{j}^{\infty} \frac{\hat{p}_{j}^2}{2m_{j}} + \frac{m_{j} \omega_{j}^2}{2}
\left(\hat{q}_{j} - \frac{c_j \hat{q}_x}{m_j\omega_j^2}\right)^2,
\end{equation}
and $\hat{H}_{\mathrm{S}} = \frac{1}{2m} \hat{p}_x ^2 + \frac{m\omega_0^2}{2} \hat{q}_x^2$, 
where $m$ and $\omega_0$ are the natural mass and frequency of the oscillator while $\hat{p}_x$
and $\hat{q}_x$  denote its 
canonical conjugate momentum and position coordinates. 
Momenta and coordinates of bath mode $j$ are denoted $\hat{p}_j$ and $\hat{q}_j$
and $c_j$ denotes the coupling term to the $j$-th mode.
In Fig.~\ref{fig:onebath}, we have depicted the physical situation.

The evolution of the system density-matrix can be analytically derived 
by means of the
Feynman-Vernon influence functional approach \cite{FV63,CL83,GSI88}. This approach
allows exploring any regime, low or high temperature, strong or weak damping, etc.
(For details on this approach  see Refs.~\cite{
FV63,CL83,GSI88,Wei08} or Ref.~\cite{PB12b}.) 
At this point, we assume that initially the density matrix of S and TB factorizes, i.e., 
$\hat{\rho}(0) = \hat{\rho}_{\mathrm{S}}(0)\otimes\hat{\rho}_{\mathrm{TB}}(0)$
and additionally that each mode of the bath is at thermal equilibrium at the same 
temperature $T_{\mathrm{TB}}$ \cite{FV63,CL83,GSI88}. 

Within this framework, the time evolution of the density matrix is given by
\begin{linenomath}
\begin{align}
\begin{split}
\langle  q_{+}'' |  & \hat{\rho}_{\mathrm{S}}(t) |  q_{-}'' \rangle =
\int \mathrm{d} q_+' \mathrm{d}q_-' 
\\ &\times J(q_+'',q_-'',t,q_+',q_-',0)
\langle q_+' | \hat{\rho}_{\mathrm{S}}(0) | q_-'  \rangle,
\end{split}
\end{align}
\end{linenomath}
where the $q$ label denotes the system coordinate representation, $J(q_+'',q_-'',t,q_+',q_-',0)$ 
is the influence functional, which is given in terms
of a path integral  after tracing out the environmental degrees of freedom
(see Refs.~\cite{FV63,CL83,GSI88,Wei08} or Ref.~\cite{PB12b}).
This solution does not provide direct insight into the dynamics in the eigenbasis of 
$\hat{H}_{\mathrm{S}}$. 
However, the solution given in  Refs.~\cite{CL83,GSI88} can be analytically transformed 
into the eigenbasis of $\hat{H}_{\mathrm{S}}$ (denoted by $\{|n\rangle\}$ with eigenvalues 
$\{E_n\}$) giving the following expression:
\begin{linenomath}
\begin{align}
\label{equ:RedDenMatEinBas}
\begin{split}
\langle n | \hat{\rho}_{\mathrm{S}}(t) | m \rangle &=
\sum_{\nu} J_{nm;\nu\nu}(t)
\langle \nu | \hat{\rho}_{\mathrm{S}}(0) | \nu \rangle
\\&+
\sum_{\nu\neq\mu} J_{nm;\nu\mu}(t)
\langle \nu | \hat{\rho}_{\mathrm{S}}(0) | \mu \rangle,
\end{split}
\end{align}
\end{linenomath}
where $J_{nm;\nu\mu}(t)$ is the influence functional in the energy basis representation,
i.e.,
\begin{equation}
\begin{split}
J_{nm;\nu\mu}(t) &= \int \mathrm{d} q_+''\mathrm{d}q_-'' \mathrm{d}q_+' \mathrm{d}q_-' 
J(q_+'',q_-'',t,q_+',q_-',0)\\
&\times \langle n| q_+''\rangle \langle q_-''| m \rangle \langle q_+'| \nu \rangle \langle \mu | q_-' \rangle.
\end{split}
\end{equation}
The result provides a remarkable linear map between the initial system state and the final 
system state that (see below) even holds in the case of initial system-bath correlations.

Of particular interest is the interplay between the diagonal elements of the density matrix 
(state populations) and the off-diagonal elements (coherences). 
In absence of coupling to the environment the kernel $J_{nm;\nu\mu}(t)$
reduces to
$J_{nm;\nu\mu}(t) = \mathrm{e}^{-\mathrm{i}(E_{m} - E_{n})t/\hbar} \delta_{n\nu}\delta_{m\mu}$.
The role of the $\delta_{n\nu}\delta_{m\beta}$ is twofold: it prevents the transfer of initial
population from $\langle \nu | \hat{\rho}_{\mathrm{S}}(0) | \nu \rangle$ to
$\langle n | \hat{\rho}_{\mathrm{S}}(t) | n \rangle$ and additionally, it eliminates any effect of initial 
off-diagonal elements on the populations (i.e., the diagonal elements). 
By contrast, in the presence of coupling to the environment, the overlap
of the system eigenstates generates ``new routes" for affecting populations,  provided
by  terms of the type $J_{nn;\nu\nu}(t)$.  
These terms  transfer the initial population 
$\langle \nu | \hat{\rho}_{\mathrm{S}}(0) | \nu \rangle$ to $\langle n | \hat{\rho}_{\mathrm{S}}(t) | n \rangle$
at time $t$. 
Additionally, any initial off-diagonal elements contribute to the time dependence  of
the populations through the nonzero $J_{nn;\nu\mu}$ terms.

For the system-bath bilinear coupling in Fig.~\ref{eq1}, the influence and nature of the 
bath is determined by a spectral density $J(\omega)$
\cite{FV63,CL83,GSI88}, which can be expressed  in terms of the bath modes
parameters:
$J(\omega) =
\pi\sum \frac{c_{j}^2}{2m_{j} \omega_{j}} \delta(\omega - \omega_{j})$. 
In this case, we assume an non-Markovian Ohmic spectral density
\begin{equation}
\label{equ:JwTB}
J(\omega) =
 m \gamma_{\mathrm{TB}}\omega
\Omega_{\mathrm{TB}}^2/\left(\Omega_{\mathrm{TB}}^2 + \omega^2\right),
\end{equation}
where $\gamma_{\mathrm{TB}}$ is the strength coupling constant to the thermal bath
and $\Omega_{\mathrm{TB}}$ is a frequency cutoff. This spectral density generates the
following damping kernel
$m \gamma(t) =
\sum_{j} \frac{c_{j}^2}{2m_{j} \omega_{j}^2} \cos(\omega_{j}t)
= 2 \int\limits_{0}^{\infty} \frac{\mathrm{d} \omega}{\pi}
           \frac{J_{\mathrm{TB}}(\omega)}{\omega}\cos(\omega t)
= \gamma_{\mathrm{TB}} \Omega_{\mathrm{TB}} \exp(-\Omega_{\mathrm{TB}}|t|)$.
This kernel is responsible for the relaxation process and describes,
roughly, the rate at which energy is transferred to the bath.
In the limit when the cutoff frequency $ \Omega_{\mathrm{TB}}$ tends to infinity,
$\gamma(t) \rightarrow 2 \gamma_{\mathrm{TB}} \delta(t)$, which corresponds
to Markovian Ohmic dissipation. 

\begin{figure}[h]
\includegraphics[width = \columnwidth]{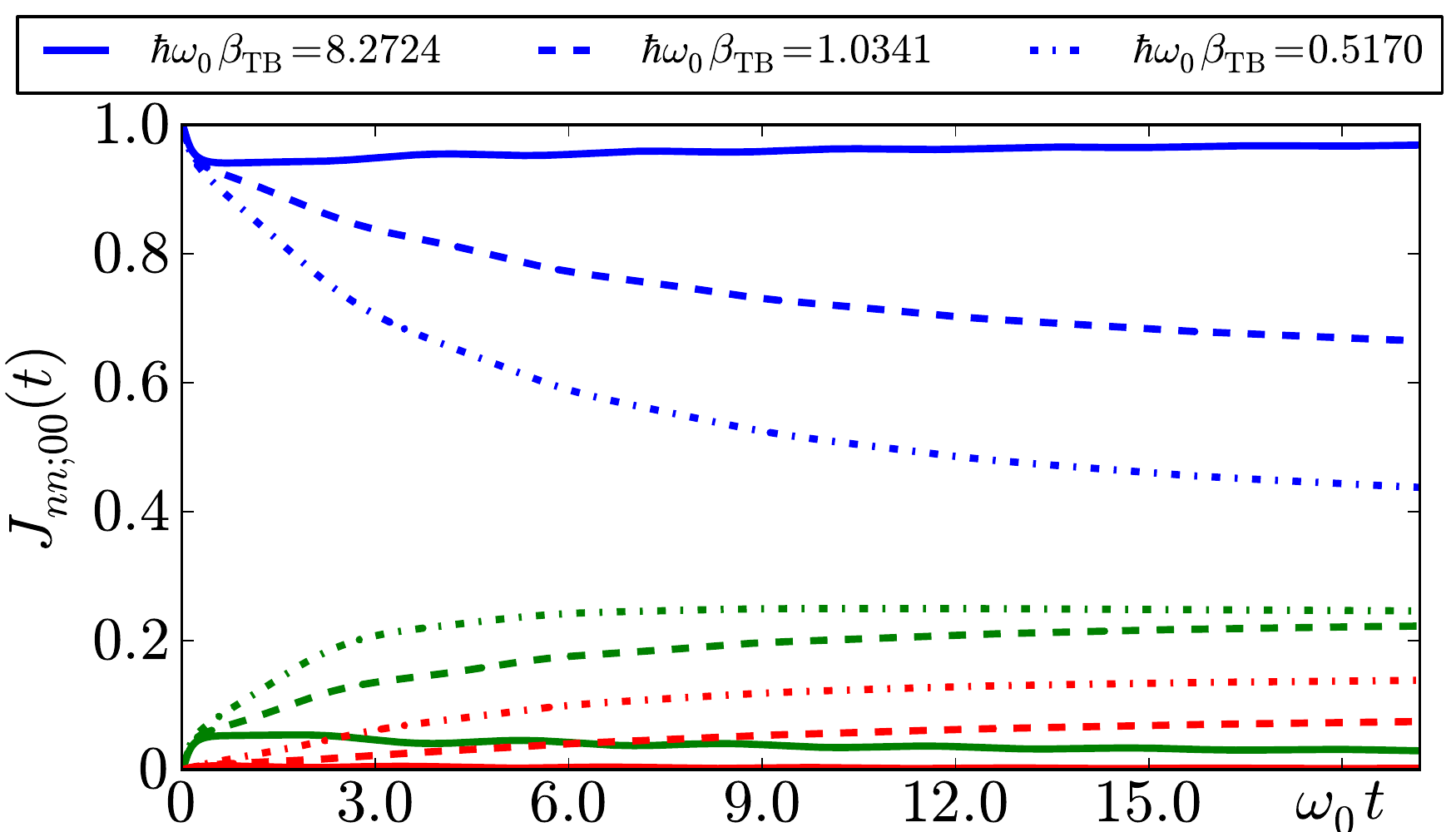}
\caption{(Color online). Time evolution of $J_{00;00}(t)$ (blue curves),
$J_{11;00}(t)$ (green curves) and
$J_{22;00}(t)$ (red curves). Results are for $\gamma_{\mathrm{TB}} = 0.1 \omega_0$,
$\Omega_{\mathrm{TB}} = 20 \omega_0$ and
$\hbar \omega_0/k_{\mathrm{B}} T_{\mathrm{TB}} = 8.2724$ (continuous curves),
$\hbar \omega_0/k_{\mathrm{B}} T_{\mathrm{TB}} = 1.0341$ (dashed curves) and
$\hbar \omega_0/k_{\mathrm{B}} T_{\mathrm{TB}} = 0.5179$ (dot-dashed curves).}
\label{fig:Jnn00}
\end{figure}

If initially $\hat{\rho}_{\mathrm{S}}(0) = \sum_n |n \rangle \langle n |$, i.e., there is  no 
initial coherence, the second sum in Eq.~(\ref{equ:RedDenMatEinBas}) vanishes and 
the time evolution of the diagonal terms depends exclusively on their initial values. 
For example, for the particular case when $\hat{\rho}_{\mathrm{S}}(0) = |0 \rangle \langle 0 |$, 
the subsequent time evolution of the diagonal elements of the density matrix can be 
expressed simply by
\begin{equation}
\label{equ:RedDenMatEinBas00}
\langle n | \hat{\rho}_{\mathrm{S}}(t) | n \rangle = J_{nn;00}(t).
\end{equation}
In Fig.~\ref{fig:Jnn00}, we have depicted $J_{nn;00}(t)$ for some values of $n$. 
By contrast to the unitary case where $J_{nn;00}(t) = \delta_{n0}\delta_{n0}$, here we have 
the possibility of populating different energy levels. 
As expected for a thermalizing system, the amount of population transfer increases with 
increasing temperature (see Fig.~\ref{fig:Jnn00}).
So, it is clear that the role of $J_{nn;00}(t)$ is transferring of population from
$\langle 0 | \hat{\rho}_{\mathrm{S}} | 0 \rangle$ to other eigenstates of S. 
This is a natural consequence of the fact that the spectrum of open quantum systems is 
broadened \cite{FV63,CL83,GSI88,Wei08}, so excitation of single energy levels is not
possible because of the overlap. 
These environmentally induced  ``new routes'' for population transfer from different eigenstates 
have been extensively exploited in the context of biological systems \cite{MR&08,PH08}. 
Here we can see that they appear mediating thermal activation, with the process being 
\emph{incoherent in nature}.

\begin{figure}[h]
\includegraphics[width = \columnwidth]{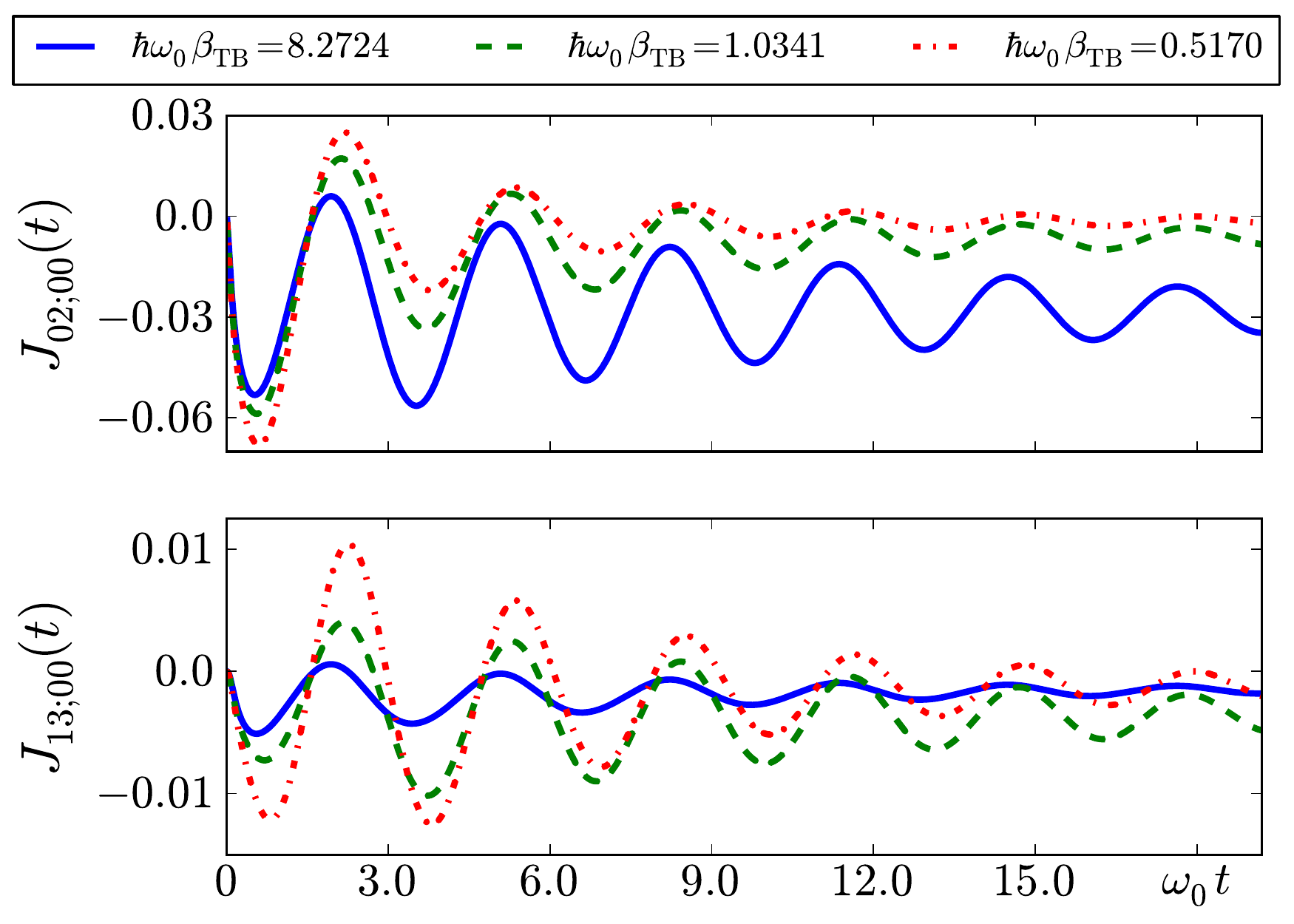}
\caption{(Color online). Time evolution of $J_{02;00}(t)$ (upper panel) and
$J_{13;00}(t)$ (lower panel). Results are for $\gamma_{\mathrm{TB}} = 0.1 \omega_0$,
$\Omega_{\mathrm{TB}} = 20 \omega_0$ and
$\hbar \omega_0/k_{\mathrm{B}} T_{\mathrm{TB}} = 8.2724$ (continuous blue curves),
$\hbar \omega_0/k_{\mathrm{B}} T_{\mathrm{TB}} = 1.0341$ (dashed green curves) and
$\hbar \omega_0/k_{\mathrm{B}} T_{\mathrm{TB}} = 0.5179$ (dot-dashed red curves).}
\label{fig:Jnm00}
\end{figure}

Remarkably, see Fig.~\ref{fig:Jnm00}, this population transfer is accompanied by the 
generation of off-diagonal terms of the density matrix, 
$\langle n | \rho_{\mathrm{S}}(t) | m \rangle = J_{nm;00}(t)$. 
By contrast to the diagonal terms, the asymptotic value of these off-diagonal terms are 
seen to increase with decreasing temperature and hence they become relevant at low 
temperature. 
For an harmonic bath, low temperature is accompanied by non-Markovian decoherence 
dynamics of the system. 
More importantly, the time decay seen in Fig.~\ref{fig:Jnm00}  is not related to 
decaying coherences because initially there was no coherence in the density matrix. 
Rather, they reflect the fact that in the \emph{effective} basis, the populations are changing 
to reach the thermal state.
It is worth noticing that these off-diagonal terms cannot be related to the existence of a 
\emph{coherent} superposition of states, but to the overlapping of energy eigenstates 
induced by the incoherent effect of the bath.

Note, significantly, that terms like $J_{nm;00}$ shown in Fig.~\ref{fig:Jnm00} do not go to zero 
at long times. 
This implies that the associated [See Eq.~(\ref{equ:RedDenMatEinBas})]  
$\langle n | \hat{\rho}_{\mathrm{S}} (t) | m \rangle$ matrix element assumes a long-time 
constant nonzero value.

In order to understand the nature of these off-diagonal terms,  note that the density
matrix is a \emph{double sided object}, which means that we can interpret $J_{nm;00}(t)$
as the influence-functional element associated with transitions from $| 0 \rangle \rightarrow | n \rangle$ 
and $\langle m | \leftarrow \langle 0 | $ mediated by thermal activation  (similar terms 
arise in optical nonlinear response \cite{Muk99}). 
At high temperature, $\hbar \omega_0 \beta_{\mathrm{TB}} \gg 1$ and 
$\frac{1}{2} \hbar \gamma_{\mathrm{TB}} \beta_{\mathrm{TB}} \gg 1$, we have
typically  ``symmetric" transitions, which are associated with transitions 
of the type $| 0 \rangle \rightarrow | n \rangle$ and $\langle n | \leftarrow \langle 0 |$.
``Asymmetric'' transitions, associated with transitions of the type 
$| 0 \rangle \rightarrow | n \rangle$ and $\langle m | \leftarrow \langle 0 |$, are less probable;
however, they are enhanced at low temperature, $\hbar \omega_0 \beta_{\mathrm{TB}} \ll 1$ and 
$\frac{1}{2} \hbar \gamma_{\mathrm{TB}} \beta_{\mathrm{TB}} \ll 1$ as seen in 
Fig.~\ref{fig:Jnm00}.  


The fact that these off-diagonal terms survive at equilibrium, points out the possibility
of deviations from ``canonical typicality" \cite{GL&06,*PSW06} (i.e., diagonal equilibrium state 
in the system energy eigenbasis) at low temperature \cite{PT13}. 
From our discussion, we can note that these off-diagonal terms are not attributable to interfering 
processes, but rather are part of thermal activation induced by the coupling to the bath, so they 
can take constant nonzero values at equilibrium rather than just vanishing. 
Based on this description, we should not call them coherences, but just static off-diagonal
terms. 
However, these have been called stationary coherences, as noted in Sec.~\ref{statcoh} and are 
simply static manifestations at equilibrium, of the system-bath coupling.

At equilibrium, the magnitude of the off-diagonal terms in the system density matrix  can be 
evaluated from the equilibrium density matrix \cite{GWT84}, i.e.,
$
\hat{\rho}_{\beta} = \mathcal{Z}_{\beta_{\mathrm{TB}}}^{-1} \mathrm{Tr}_{\mathrm{TB}}
\exp\left[-\beta_{\mathrm{TB}}(\hat{H}_{\mathrm{S}} + \hat{H}_{\mathrm{ST}} 
+ \hat{H}_{\mathrm{TB}})\right]
$, 
where $\mathcal{Z}_{\beta_{\mathrm{TB}}}$ is a normalization factor and Tr$_{\mathrm{TB}}$ denotes 
the trace over the bath. 
For this particular case, we can introduce the effective Hamiltonian
$
\hat{H}_{\mathrm{eff}} = \frac{1}{2 m_{\mathrm{eff}}} \hat{p}_x^2
+ \frac{1}{2}m_{\mathrm{eff}} \omega_{\mathrm{eff}}^2 \hat{q}_x^2,
$
with the effective mass
$
m_{\mathrm{eff}} =
\omega_{\mathrm{eff}}^{-1} \sqrt{\langle p^2\rangle\langle q^2\rangle^{-1}}, $
 and the effective frequency 
$
\omega_{\mathrm{eff}} = 2(\hbar \beta_{\mathrm{TB}})^{-1} \mathrm{arccoth}\left(
\frac{2}{\hbar}\sqrt{\langle p^2\rangle\langle q^2\rangle}\right),
$ 
while $\langle q^2\rangle$ and $\langle p^2\rangle$ being the equilibrium variances
\cite{GWT84,GSI88}. This allows expressing $\hat{\rho}_{\beta}$ as \cite{GWT84}
\begin{equation}
\hat{\rho}_{\beta} = Z_{\beta_{\mathrm{TB}}}^{-1} \sum_{n=0}^{\infty}
 \exp(-E_{n_{\beta}} \beta_{\mathrm{TB}}) | n_{\beta} \rangle \langle n_{\beta} |,
\end{equation}
with $E_{n_{\beta}} = \hbar \omega_{\mathrm{eff}}(n_\beta + \frac{1}{2})$  being the 
eigenvalues
and $| n_{\beta} \rangle$ the eigenstates of the \textit{effective Hamiltonian} $\hat{H}_{\mathrm{eff}}$
and $Z_{\beta_{\mathrm{TB}}}$ is the generalized partition function \cite{GWT84}.
At high temperature, $\hbar \omega_0 \beta_{\mathrm{TB}} \ll 1$
and $\frac{1}{2} \hbar \gamma_{\mathrm{TB}} \beta_{\mathrm{TB}} \ll 1$,
$m_{\mathrm{eff}}$ and $\omega_{\mathrm{eff}}$ approaches their bare values
$m$ and $\omega_0$, respectively, and $\hat{\rho}_{\beta}$ tends to the canonical
distribution  \cite{GWT84,HI05} (with no off-diagonal terms in the system energy eigenbasis). 
At low temperatures,
$\hbar \omega_0 \beta_{\mathrm{TB}} \gg 1$ and 
$\frac{1}{2} \hbar \gamma_{\mathrm{TB}} \beta_{\mathrm{TB}} \gg 1$, they undergo
to strong deviation due to damping \cite{GWT84,HI05}.

Since we are interested in natural processes, we next consider excitation of the equilibrated 
system S by (\emph{i}) a second thermal bath TB$^\prime$ of the same nature as TB 
but at a different temperature and different coupling constant, and (\emph{ii}) by blackbody 
radiation, denoted BB.  
For example, in the particular case of biological processes, e.g., in electronic energy 
transfer
in photosynthetic complexes \cite{EC&07,CW&10},  the coupling to the environment is strong
\cite{CF09} and the effective temperature is low \cite{PB11,PB12}. 
So, a question immediately follows:  do these off-diagonal terms play any relevant role in any 
subsequent  dynamics of these systems? 
More specifically, in what way do they contribute to any dynamics arising from subsequent 
perturbation?

\section{Dynamics in the presence of a second thermal bath}
\label{TBPcase}
\begin{figure}
\includegraphics[width = 0.75\columnwidth]{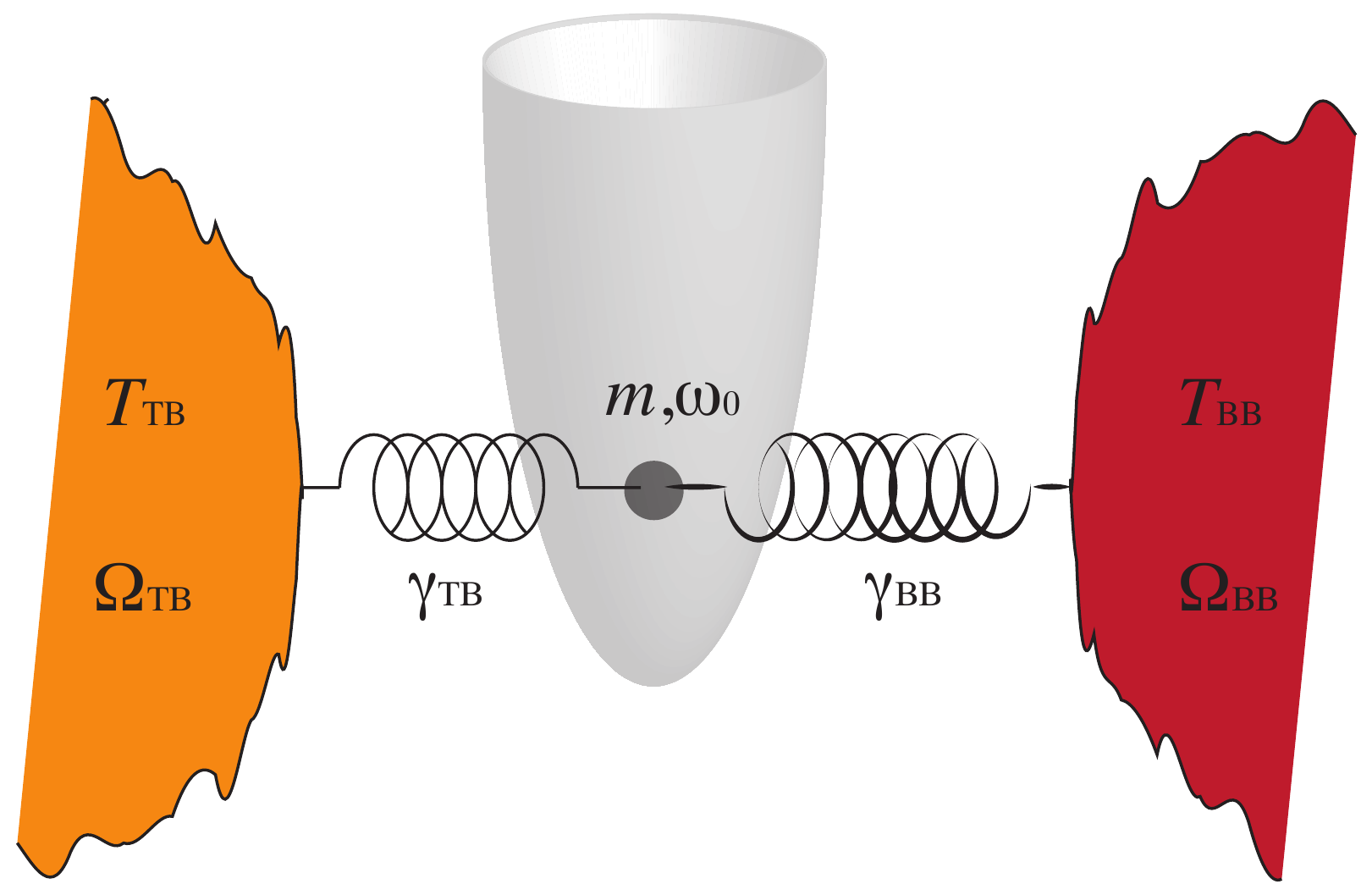}
\caption{After thermalizing with TB, the system S is put in contact with a second thermal 
bath TB$^\prime$ (or BB, as in the plot) at different temperature and different coupling 
constant.}
\label{fig:twobaths}
\end{figure}

Equilibration of a system at temperature $T_{\mathrm{TB}}$, with another bath at temperature 
$T_{\mathrm{TB}^\prime}$ is of
 general interest.  
 For example, for biological systems, thermal
activation of biological processes by temperature changes can be found, e.g., in the
context of transport of Ca ions through membranes \cite{XC&11}.  
To examine such processes, after thermalizing with TB, we couple the oscillator to a second
dissipative environment TB$^\prime$. The Hamiltonian of the system can now be written as
\begin{linenomath}
\begin{equation}
\label{equ:HamSys}
\hat{H} = \hat{H}_{\mathrm{S}} + \hat{H}_{\mathrm{TB}} + \hat{H}_{\mathrm{TB}^\prime}
   + \hat{H}_{\mathrm{ST}} + \hat{H}_{\mathrm{ST}^\prime},
\end{equation}
\end{linenomath}
where \(\hat{H}_{\mathrm{TB}^\prime}\) is the Hamiltonian describing the second thermal bath 
TB$^\prime$, while \(\hat{H}_{\mathrm{ST}^\prime}\) describes the interaction of the system 
with TB$^\prime$. 
As in the previous case, we choose \(\hat{H}_{\mathrm{TB}^\prime}\) as being composed of a 
set of harmonic oscillators, so
\(
\hat{H} = \hat{H}_{\mathrm{S}}
  + \sum_{j,k}^{\infty,2} \frac{\hat{p}_{j,k}^2}{2m_{j,k}}
  + \frac{m_{j,k} \omega_{j,k}^2}{2}  \left(\hat{q}_{j,k} - \frac{c_{j,k} \hat{q}_x}{m_{j.k} \omega_{j,k}^2}\right)^2.
\)
Note that the coupling is of the system S to the heat bath TB and of the system to the 
heat bath TB$^\prime$, as shown in Fig.~\ref{fig:twobaths}. There is no direct coupling  
between TB and TB$^\prime$.

The resultant evolution of the system density-matrix can also be analytically obtained  
by using the influence functional approach \cite{PB12b}. 
The evolution is of the form in Eq.~\ref{equ:RedDenMatEinBas}, 
but with $\hat{\rho}_{\mathrm{S}}(0) = \hat{\rho}_{\beta}$ and the $J_{nm;\nu\mu}(t)$'s
containing information about the effect of the initial correlation on the subsequent dynamics. 

In our approach, we have \emph{exact analytical} access to \emph{every contribution} 
to the dynamics. 
By contrast, for example,  when applying the secular approximation as in Ref.~\cite{MV10}, 
the contribution from the off-diagonal terms $\nu\neq\mu$ is ignored.

It is important to note that in this case the initial condition is 
$\hat{\rho}(0) = \hat{\rho}_{\mathrm{S}+\mathrm{TB}}(0)\otimes\hat{\rho}_{\mathrm{TB}^\prime}(0)$,
where $\hat{\rho}_{\mathrm{S}+\mathrm{TB}}(0)$ is the equilibrium density operator of 
(S$+$TB), and  $\hat{\rho}_{\mathrm{TB}^\prime}(0)$ denotes the thermal density operator 
of the second bath at temperature $T_{\mathrm{TB}^\prime}$. 
We describe the effect of the baths using, for both TB and TB$^\prime$, the functional form 
given in Eq.~(\ref{equ:JwTB}).
In the absence of TB$^\prime$,  the overall (S$+$TB) is time-independent, as it should be 
thermal equilibrium \cite{SG86,GSI88,PID10}.

During this second equilibration step, the off-diagonal elements of the system density 
matrix that were generated during the thermalization step affect  the population dynamics 
(the diagonal terms) because  they 
now enter in the initial density matrix [see Eq.~(\ref{equ:RedDenMatEinBas})].  

\begin{figure}[h]
\includegraphics[width = \columnwidth]{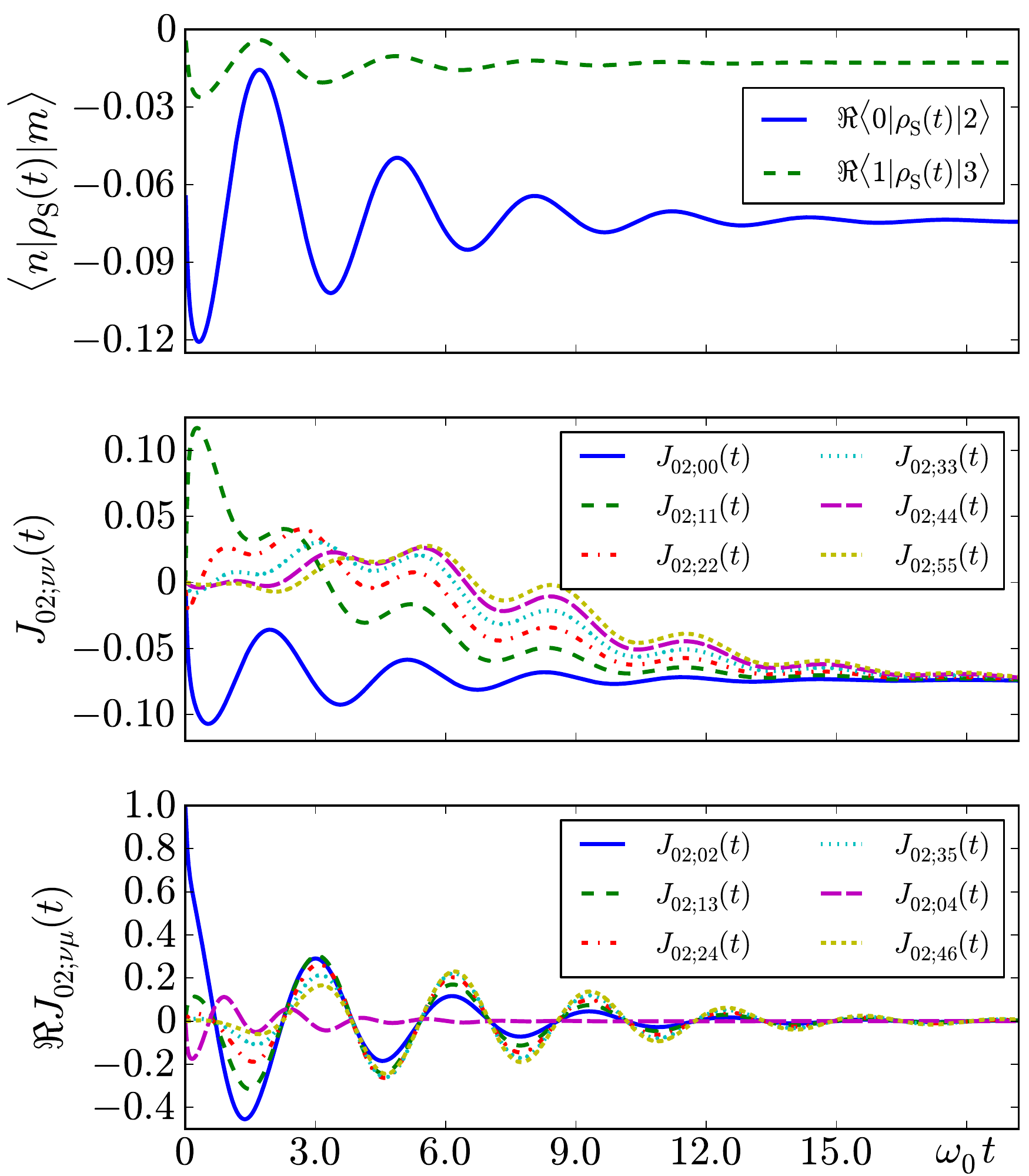}
\caption{(Color online). Upper panel: exact time evolution of $\langle 0|\hat{\rho}_{\mathrm S}|2\rangle$ and
$\langle 1|\hat{\rho}_{\mathrm S}|3\rangle$ with 
$\hbar \omega_0/k_{\mathrm{B}} T_{\mathrm{TB}} = 8.2724$,
$\gamma_{\mathrm{TB}} = 0.1 \omega_0$, $\Omega_{\mathrm{TB}^\prime} = 20 \omega_0$,
$T_{\mathrm{TB}^\prime} =2 T_{\mathrm{TB}}$,
$\gamma_{\mathrm{TB}} = 2 \gamma_{\mathrm{TB}}$ and $\Omega_{\mathrm{TB}^\prime}
= 2\Omega_{\mathrm{TB}}$. Lower panels: time evolution of $J_{02;\nu\mu}$ for
some values of $\nu$ and $\mu$.}
\label{fig:J02nnTBTB}
\end{figure}
In the upper panel of Fig.~\ref{fig:J02nnTBTB}, we present the time evolution of
$\langle 0|\hat{\rho}_{\mathrm S}(t)|2\rangle$ and $\langle 1|\hat{\rho}_{\mathrm S}(t)|3\rangle$
for the conditions indicated in the figure caption. 
The off-diagonal terms are seen to reach the same order of magnitude as the populations 
(not shown), i.e. $10^{-1}$. 
Time evolution of, e.g., $\langle 0|\hat{\rho}_{\mathrm S}(t)|2\rangle$ is affected  by terms of the 
type $J_{02,\nu\nu}$ and $J_{02,\nu\mu}$. 
In particular,  population transfer is assisted by $J_{02,\nu\nu}$ and decay of the initial coherences 
is mediate by $J_{02,\nu\mu}$.
In the central and lower panels of Fig.~\ref{fig:J02nnTBTB}, we have depicted $J_{02,\nu\nu}$
and $J_{02,\nu\mu}$, respectively, for various values of $\nu$ and $\mu$. 
There we can see that while  $J_{02,\nu\nu}$ reach an asymptotic finite value,  $J_{02,\nu\mu}$ 
goes to zero as time evolves. 
Hence, it is clear that terms like $J_{02,\nu\mu}$ are qualitatively different  from $J_{02,\nu\nu}$ 
type terms. 
The former are related to the decay of quantum coherences while the latter are related to population 
transfer during the equilibration.

\begin{figure}[h]
\includegraphics[width = \columnwidth]{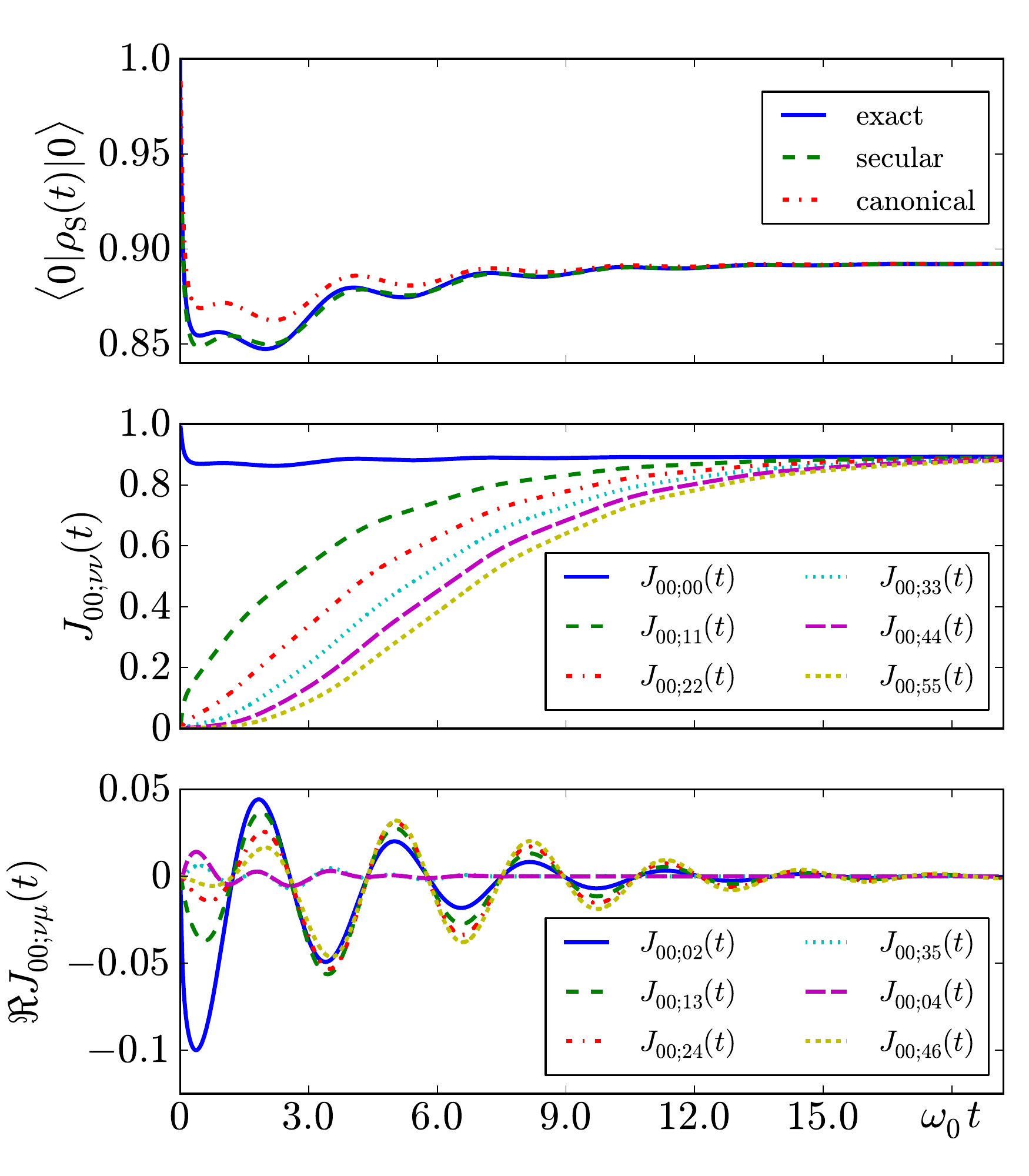}
\caption{(Color online). Upper panel: exact (continuous blue curve), secular-approximated (dashed
green curve) time evolution of $\langle 0|\hat{\rho}_{\mathrm S}|0\rangle$. 
The evolution using the canonical distribution as the initial state is depicted by the dot-dashed 
red curve. 
Lower panels: time evolution $J_{00;\nu\mu}$ for some values of $\nu$ and $\mu$. 
Parameters as in Fig.~\ref{fig:J02nnTBTB}.}
\label{fig:TBTB}
\end{figure}

Figure \ref{fig:TBTB} shows the time evolution of the ground state population 
 $\langle 0|\hat{\rho}_{\mathrm S}(t)|0\rangle$.  
The exact evolution is depicted by the continuous blue curve, the evolution disregarding 
any correlation with TB [i.e., using the canonical distribution for $\hat{\rho}_{\mathrm{S}}(0)$] 
is shown by the dot-dashed red curve and the evolution neglecting the off-diagonal terms 
(i.e. the secular approximation) in Eq.~(\ref{equ:RedDenMatEinBas}) is shown by the dashed 
green curve. 
The fact that the dynamics in these three cases is practically the same shows  that, although 
off-diagonal terms are present in Fig.~\ref{fig:J02nnTBTB}, they need not imply a significant 
contribution to the dynamics between eigenstates.
That is, the presence of off-diagonal terms in the density matrix of open quantum systems 
in the system eigenstate representation is not necessarily  an indicator of coherent effects.  
The lower panels in Fig.~ \ref{fig:TBTB} show some of the non-vanishing $J_{00;\nu\mu}(t)$ elements.

It is worth noting that off-diagonal terms generated by the presence of the second bath
cannot manipulate the dynamics of the populations by themselves. This can be seen
when projecting in the basis in which $\hat{\rho}_{\beta}$ is diagonal, i.e.
$
\langle n_{\mathrm{\beta}} | \hat{\rho}_{\mathrm{S}}(t) | n_{\mathrm{\beta}} \rangle =
\sum_{\nu_{\mathrm{\beta}}}
J_{n_{\mathrm{\beta}}n_{\mathrm{\beta}};\nu_{\mathrm{\beta}}\nu_{\mathrm{\beta}}}(t)
\langle \nu_{\mathrm{\beta}} | \hat{\rho}_{\beta} | \nu_{\mathrm{\beta}} \rangle$.
This means that although off-diagonal terms are generated (see Fig.~\ref{fig:J02nnTBTB}),
they do not participate in the evolution of the populations. 
Only the combined action of the thermal baths, applied sequentially, can lead to the possibility 
of altering the dynamics of the populations. 
Connecting the system to the baths, either \emph{separately} or \emph{simultaneously} 
only leads to  excitation of the system.

\section{Dynamics Induced by Blackbody Radiation}
\label{BBcase}
Consider now the case where the second thermal bath TB$^\prime$ is replaced by blackbody 
radiation (BB).  Here,  a charged harmonic oscillator is immersed in a dissipative environment 
TB and coupled via dipole approximation to blackbody radiation BB. 
This provides a generic model for  a wide variety of objects such as atoms, ions, electrons, 
molecules in equilibrium, subjected to blackbody irradiation. 
Qualitative results for this case are particularly relevant for natural light incident on biomolecules.

The Hamiltonian of the total system is now of the form of Eq.~(\ref{equ:HamSys}) where
$\hat{H}_{\mathrm{B}}$ is the Hamiltonian describing the radiation field and 
$\hat{H}_{\mathrm{SB}^\prime}$ describes the interaction of the system with the blackbody 
radiation. 
In the dipole approximation, the Hamiltonian for an oscillator interacting with a radiation 
field and coupled linearly to its surrounding environment is
\begin{linenomath}
\begin{equation}
\begin{split}
\label{equ:HamSysBB}
&\hat{H} = \frac{1}{2m}\left(\hat{p}_x - \frac{e}{c} \hat{A}_x\right)^2
   + \frac{m\omega_0^2}{2} \hat{q}_x^2+ \sum_{j} \left[\frac{\hat{p}_j^2}{2m_j} \right.
   \\   & 
 \left. + \frac{m_j \omega_j^2}{2}  (\hat{q}_j - \frac{c_j \hat{q}_x}{m_j \omega_j^2})^2\right]
   + \sum_{\mathbf{k},s}\hbar c k
   \left(\hat{a}_{\mathbf{k},s}^{\dagger} \hat{a}_{\mathbf{k},s}+\frac{1}{2}\right) ,
\end{split}
\end{equation}
\end{linenomath}
where $e/c$ is the coupling constant to the radiation, $\hat{a}_{\mathbf{k},s}$ and
$\hat{a}_{\mathbf{k},s}^{\dagger}$ are the annihilation and creation operators of the
field mode of momentum $\mathbf{k}$ and polarization $s$. 
The vector potential is given by
\begin{align}
\label{equ:AVPBBR}
 \hspace{-0.175cm} \hat{A}_x = \sum_{\mathbf{k},s} \hspace{-0.05cm} \left[\frac{h c}{kV}\right]^{\frac{1}{2}}
 \hspace{-0.1cm}\left[
     f_k^* \hat{a}_{\mathbf{k},s}\mathbf{e}_{\mathbf{k},s}\cdot \mathbf{q}
    +f_k \hat{a}_{\mathbf{k},s}^{\dagger}\mathbf{e}_{\mathbf{k},s}^*\cdot \mathbf{q}
    \right],
\end{align}
where $\mathbf{e}$ is the polarization vector, $V$ is the volume of the cavity and
$f_k$ is the electron form-factor (Fourier transform of the charge distribution) which
incorporates the electron structure \cite{FLO85,*FLO88}.

By defining  $m_{\mathbf{k}} = 4\pi e^2f_k^2/(\omega_{\mathbf{k}} V)$ and
$a_{\mathbf{k},s} = (m_{\mathbf{k}} \omega_{\mathbf{k}} q_{\mathbf{k},s}
+ \mathrm{i}p_{\mathbf{k},s})/\sqrt{2m_{\mathbf{k}} \hbar \omega_{\mathbf{k}}}$, and
applying the Power-Zienau's transformation (see Ref.~\cite{PB12b} for details)
we can rewrite Eq.~(\ref{equ:HamSysBB}) as
\begin{linenomath}
\begin{equation}
\label{equ:FinalHamBB}
\begin{split}
\hat{H} &= \frac{1}{2m} \hat{p}_x^2
   + \frac{1}{2}m\omega_0^2 \hat{q}_x^2
\\ &+ \sum_{\mathbf{k},s}
   \frac{1}{2 m_{\mathbf{k}}}\left(\hat{p}_{\mathbf{k},s} +
   m_{\mathbf{k}} \omega_{\mathbf{k}} \hat{q}_x \right)^2 +
   \frac{1}{2}m_{\mathbf{k}}\omega_{\mathbf{k},s}^2 \hat{q}_{\mathbf{k},s}^2
\\&+\sum_{j}\frac{\hat{p}_j^2}{2m_j}  + \frac{1}{2} m_j \omega_j^2 \left(\hat{q}_j 
- \frac{c_j \hat{q}_x}{m_j \omega_j^2}\right)^2,
\end{split}
\end{equation}
\end{linenomath}
where the oscillator is seen to be coupled to the momentum coordinate
$\hat{p}_{\mathbf{k},s}$. 
Since $\hat{E}_x = - \partial \hat{A}_x /\partial t$ and based on the definition of the annihilation 
and creation operators (see above), one can show that this momentum coupling is equivalent 
to the oscillator being coupled to the electric field of the radiation with
\begin{linenomath}
\begin{equation*}
\hat{E}_x = \textrm{i} \sum_{\mathbf{k},s} \left(\frac{h c^3}{V}\right)^{\frac{1}{2}}
     \left(
     f_k^* \hat{a}_{\mathbf{k},s}\mathbf{e}_{\mathbf{k},s}\cdot \mathbf{q}
    +f_k \hat{a}_{\mathbf{k},s}^{\dagger}\mathbf{e}_{\mathbf{k},s}^*\cdot \mathbf{q}
    \right).
\end{equation*}
\end{linenomath}

In order to describe the action of the blackbody radiation, 
we need to consider, as in
the previous case, that the modes in the cavity are thermally populated (for more details 
see Ref.~\cite{PB12b}), i.e., each mode is characterized by an incoherent 
density operator.
This feature introduces the incoherent nature of the radiation 
considered in this work.
Based on this description, we expect that the blackbody radiation generates an 
incoherent evolution in contrast with the coherent evolution induced by laser pulses.

Interestingly, the thermal fluctuations generated by the blackbody radiation are characterized 
by a two--point electric--field correlation function  $\langle \hat{E}_x(t'') \hat{E}_x(t')\rangle$ 
that is  not $\delta$-correlated. 
This reveals the intrinsic non-Markovian character of the radiation
from both a statistical viewpoint \cite{FLO85,*FLO87,*FLO88,BC91,PB12b} and from the
optics point of view \cite{MW64a,*MW64b,*MW67}. 
The correlation time $\tau_{\mathrm{BB}}^{\mathrm{c}}$ of the randomly fluctuating electric 
field can be calculated from the thermal time
$\tau_{\mathrm{BB}}^{\mathrm{th}} =\hbar/ k_{\mathrm{B}} T_{\mathrm{BB}}$  \cite{MW64a,
*MW64b,*MW67} and is expressed roughly as
$\tau_{\mathrm{BB}}^{\mathrm{c}} \sim \tau_{\mathrm{BB}}^{\mathrm{th}}$.
For blackbody radiation at
\(T_{\mathrm{BB}} = 300~\)K, $\tau_{\mathrm{BB}}^{\mathrm{c}} \sim 25.5~$fs whereas for
sunlight, \(T_{\mathrm{BB}} = 5900~\)K,   $\tau_{\mathrm{BB}}^{\mathrm{c}} \sim 1.3~$fs
and for moonlight,
\(T_{\mathrm{BB}} = 4100~\)K,   $\tau_{\mathrm{BB}}^{\mathrm{c}} \sim 1.86~$fs.
Thus, for processes taking place on the order of, e.g., 1~ps (such as electronic energy
transfer in photosynthetic complexes \cite{EC&07,CW&10}) this coherence time
is very short.  Hence, under illumination by sunlight, the perturbation is effectively CW and 
incoherent.

From an open-quantum-system perspective, the influence of the blackbody radiation 
on the system is condensed in the spectral density
\cite{FLO85,*FLO87,*FLO88,PB12b,BC91}
\begin{equation}
\label{equ:JwBB}
J_{\mathrm{BB}}(\omega) = M \tau_{\mathrm{BB}} \, \omega^3 \Omega_{\mathrm{BB}}^2/
\left(\Omega_{\mathrm{BB}}^2 + \omega^2\right),
\end{equation}
where $M = m + M \tau_{\mathrm{BB}} \Omega_{\mathrm{BB}}$ is the renormalized mass,
$\tau_{\mathrm{BB}} = 2 e^2 /3 M_\mathrm{e} c^3 \sim 6.24\times10^{-24}$s
and $\Omega_{\mathrm{BB}}$ is a frequency cutoff \cite{FLO85,*FLO87,*FLO88}.
This spectral density generates the following dissipative kernel
$
\gamma_{\mathrm{BB}}(t) =\tau_{\mathrm{BB}} \Omega_{\mathrm{BB}}^2
\left[2 \delta(t) - \Omega_{\mathrm{BB}} \exp(-\Omega_{\mathrm{BB}}|t|)\right].
$
Note that there is a fundamental limitation to the use of Eq.~(\ref{equ:JwBB}).
That is, in the limit $\Omega_{\mathrm{BB}} \rightarrow\infty$, we get the 
surprising result,
$\gamma_{\mathrm{BB}}(t) = 0$, i.e. no relaxation. 
This corresponds to  the point-electron limit
[$f_k^2 = \Omega_{\mathrm{BB}}^2/(\Omega_{\mathrm{BB}}^2 + \omega_k^2) = 1$ in
Eq.~(\ref{equ:AVPBBR})] and is unphysical
 because
even for the electron, $\Omega_{\mathrm{BB}}$ remains finite, although large.
There is a natural upper limit given by \cite{FO91,*OCo03,FO98}
$\Omega_{\mathrm{BB}} = {\tau_{\mathrm{BB}}}^{-1}$,
which corresponds to 2/3 of the time for photon to traverse the classical electron radius
($r_\mathrm{cl}^{\mathrm{e}} = 2.818\times10^{-15}$m). Beyond this natural limit,
causality is violated \cite{FO91,*OCo03} and the bare mass $m$ takes negatives values 
\cite{FO91,*OCo03}.
Note that one could consider another reasonable choice of form factor, e.g., 
$f_k^2 = \Omega_{\mathrm{BB}}^4/(\Omega_{\mathrm{BB}}^2 + \omega_k^2)^2$, corresponding
to a sharper cut-off \cite{FO91,*OCo03}. 
This will lead to corrections in the equation of motion of the order of $\tau_{\mathrm{BB}}$
and $\tau_{\mathrm{BB}}^2$, which compared with $\tau_{\mathrm{BB}}^{\mathrm{c}}$ 
are negligible. 
Thus, following \cite{FO91,*OCo03}, 
$f_k^2 = \Omega_{\mathrm{BB}}^2/(\Omega_{\mathrm{BB}}^2 + \omega_k^2)$ can 
be considered as an excellent approximation.

The exact analytic expression for the influence functional for this case is derived in 
Ref.~\cite{PB12b}. 
There we show that the equations of motion are driven by a \emph{transient term} that 
is dependent on the initial conditions and proportional to $\gamma_{\mathrm{BB}}$.
This term is absent in the former case above, and is a consequence of the coupling 
to the momentum of the modes rather than coupling through the coordinate of the
modes [cf. Eq.~(\ref{equ:FinalHamBB})]. 
So, in addition to the turn-on effect, present in the former case, in the case of 
incoherent excitation by blackbody radiation, we also have a driven transient term.

In order to gain insight into the strength of this transient term and of the blackbody radiation, 
we examine results in the limit when $\Omega_{\mathrm{BB}} \rightarrow \tau_{\mathrm{BB}}^{-1}$, 
and where the effect of the radiation can be estimated by a constant damping kernel given 
by $\gamma_{\mathrm{BB}} = \omega_0^2 \tau_{\mathrm{BB}}$ \cite{FLO85}. 
Assuming a typical value of electronic molecular structure (e.g. a carbon-carbon bound) 
$\omega_0 = 3 \times 10^{14}$~Hz, we have that 
$\gamma_{\mathrm{BB}} = 1.8 \times 10^{-9}\omega_0$. 
In this case the excitation due to the radiation is too weak to  compete with the incoherent 
effect of TB and, in absence of TB, the thermalization would take far too long 
($\gamma_{\mathrm{BB}}^{-1} = 1.85$~$\mu$s) to be appreciable on, e.g., a picosecond 
time-scale. 
In order to see some sort of appreciable effect on the picoseconds time scale would
require a cutoff on the order of 
$\Omega_{\mathrm{BB}}  = 5\times10^{-6} \tau_{\mathrm{BB}}^{-1}$
($\gamma_{\mathrm{BB}}^{-1} =9$~ps).

\begin{figure}[h]
\includegraphics[width = \columnwidth]{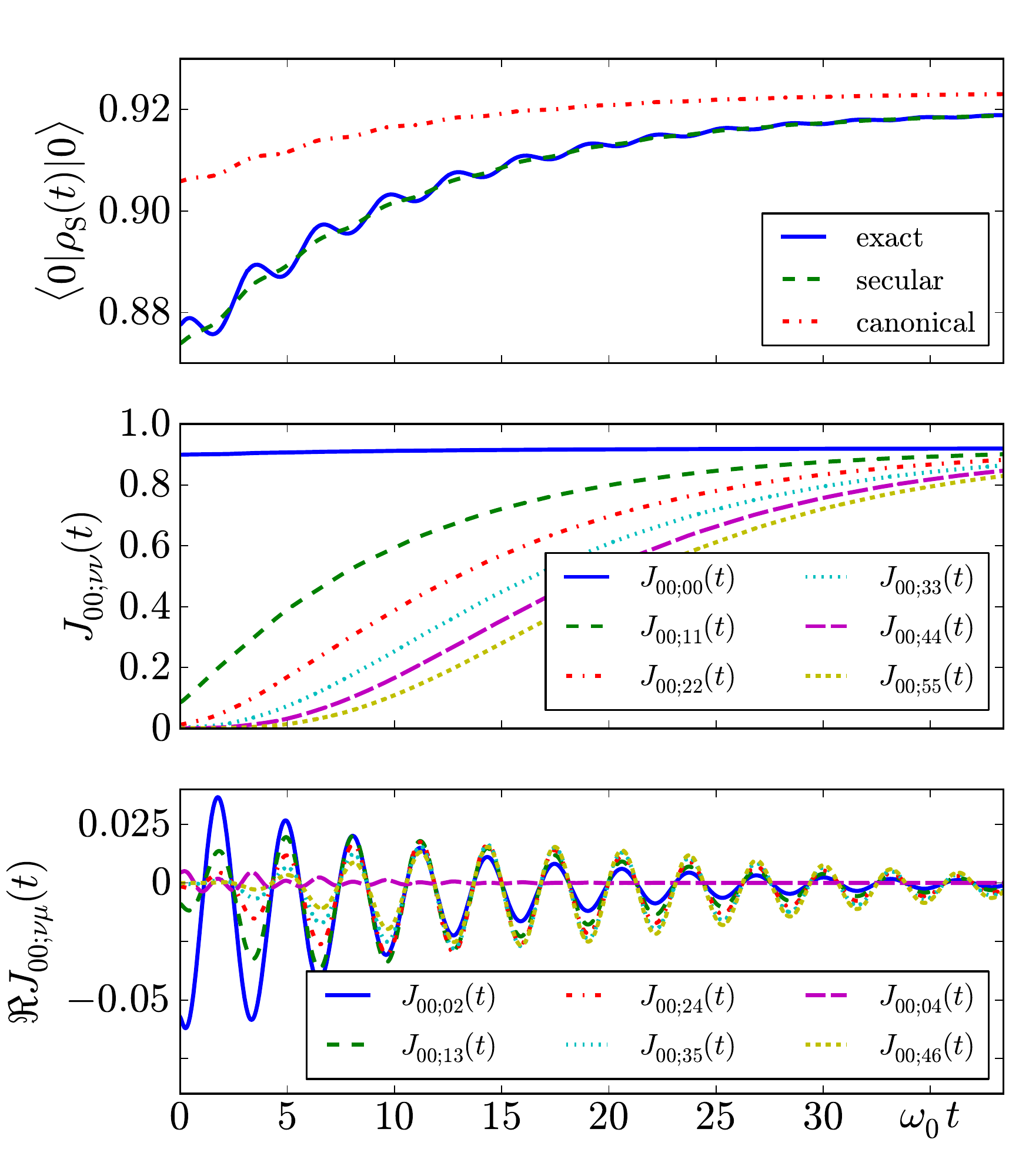}
\caption{(Color online). Upper panel: exact (continuous blue curve), secular-approximated (dashed
green curve) time evolution of $\langle 0|\hat{\rho}_{\mathrm S}|0\rangle$. 
The evolution using the canonical distribution as the initial state is depicted by the dot-dashed 
red curve. 
Lower panels: time evolution $J_{00;\nu\mu}$ for some values of $\nu$ and $\mu$. 
Parameters for coupling to TB as in Fig.~\ref{fig:TBTB}, with 
$\hbar \omega_0/k_{\mathrm{B}} T_{\mathrm{BB}} = 0.3884$,
and $\Omega_{\mathrm{BB}} = 8.3 \times 10^{5}\omega_0$.}
\label{fig:TBBB}
\end{figure}

\begin{figure}[h]
\includegraphics[width = \columnwidth]{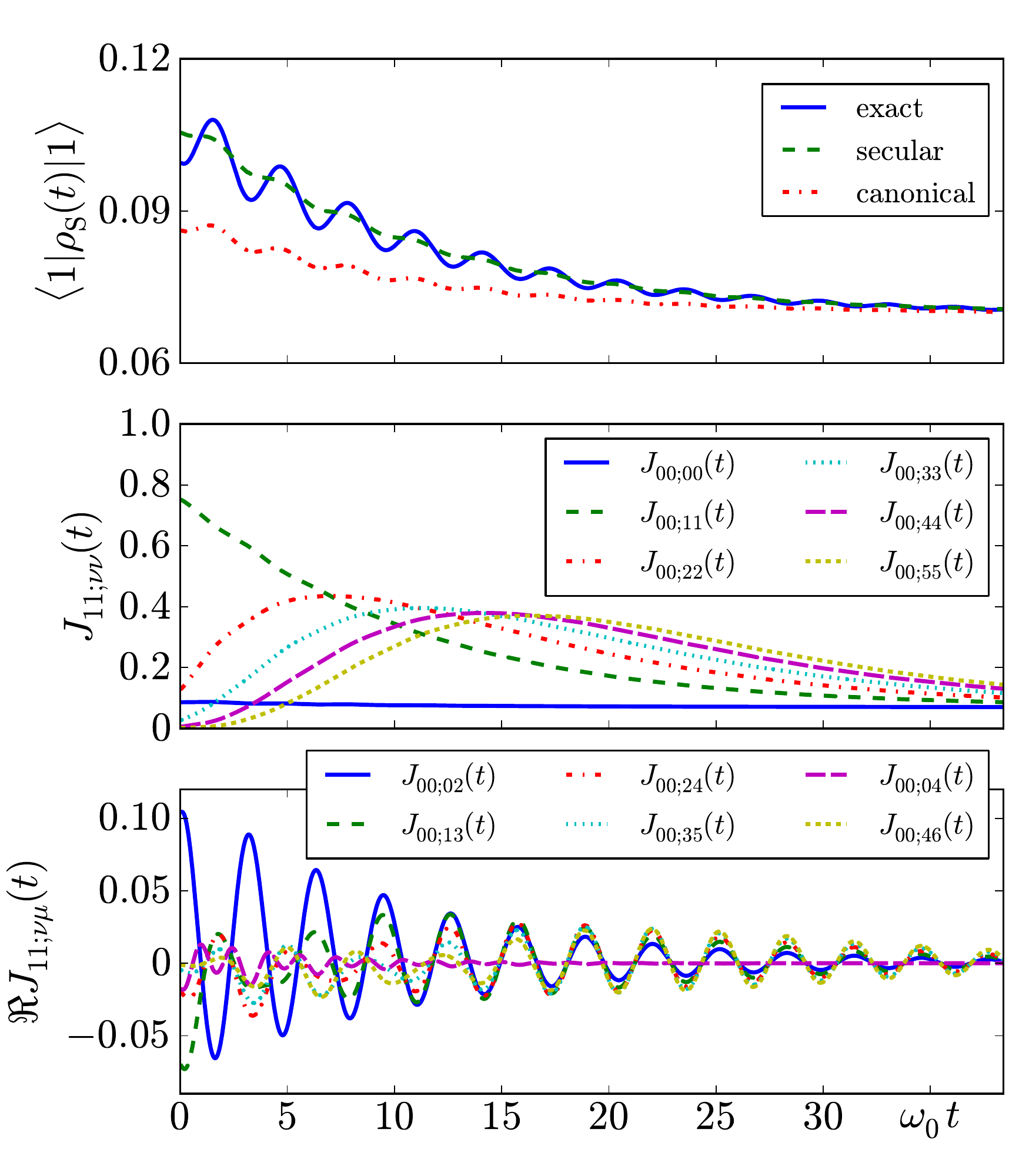}
\caption{(Color online). Time evolution of  $\langle 1|\hat{\rho}_{\mathrm S}|1\rangle$. 
Description and parameters as in Fig.~\ref{fig:TBBB}}
\label{fig:rho11TBBB}
\end{figure}

As an example of the dynamics, Fig.~\ref{fig:TBBB} shows the time evolution of the ground 
state and some propagating elements $J_{00,\nu\mu}(t)$ and Fig.~\ref{fig:rho11TBBB} 
provides  the corresponding results for the first excited state.
In order to see some sort of appreciable effect, we have used here the artificial value 
of $\Omega_{\mathrm{BB}}  = 5\times10^{-6} \tau_{\mathrm{BB}}^{-1}$ and additionally 
strong coupling to the environment $\gamma_{\mathrm{TB}} = 10^{-1}\omega_0$.  
With these parameters, we have that in Fig.~\ref{fig:TBBB} and Fig.~\ref{fig:rho11TBBB}, 
$T_{\mathrm{TB}} = 277$~K while $T_{\mathrm{BB}} = 5900$~K. 
For this frequency, the time interval depicted in Fig.~\ref{fig:TBBB} and Fig.~\ref{fig:rho11TBBB} 
is $\approx$ 128~fs.
Here, the exact evolution of $\langle 0|\hat{\rho}_{\mathrm{S}}(t)|0\rangle$ and  
$\langle 1|\hat{\rho}_{\mathrm{S}}(t)|1\rangle$ are depicted by a continuous blue line, 
the time evolution neglecting the off-diagonal (stationary coherent) terms initially 
generated by the presence of the TB is depicted using the dashed green line, and the 
evolution using the canonical distribution as the initial state by the dot-dashed red curve. 

In the lower panels, we present some of the non-vanishing $J_{00;\nu\mu}(t)$ and 
$J_{11;\nu\mu}(t)$ elements. 
It is evident that the effect of the radiation is so weak that it generates neither off-diagonal 
terms nor significant changes in the populations dynamics. 
That is, even with the choice of excessively aggressive parameters, the effect of the 
radiation is negligible. 
Rather, the time evolution can be seen as being caused by the turn-on of the interaction 
followed by the subsequent relaxation of the system to equilibrium, dominated by the 
interaction with TB.
The weak oscillations in Fig.~\ref{fig:TBBB} and Fig.~\ref{fig:rho11TBBB}, can be attributed 
to the transient driven term, which is strongly determined by the frequency cutoff 
$\Omega_{\mathrm{BB}}$. 
For different sets of parameters these oscillations might well be absent.

The effect of switching the interaction on $t=0$ and the time-dependent 
driving transient term can be also appreciated by the jump of the $J_{nm;\nu\mu}$-terms. 
That is, in Fig.~\ref{fig:TBBB},  $J_{00;00}$ is seen to jump from unity at $t=0$
to a different value, and $J_{00;11}$ and $J_{00;\nu\mu}$ jump from zero to a finite
value. The same applies for $J_{11;00}$ and $J_{11;\nu\mu}$ in Fig.~\ref{fig:rho11TBBB}.

From  Fig.~\ref{fig:TBBB}, we can see oscillations in the population of the ground  state,  
these oscillations are also present in, e.g., the  population  of  the  first  excited  state  [see 
Fig.~\ref{fig:rho11TBBB}]. 
Since,  in  our ``secular approximation" these oscillations are removed,  according to
us, this  would imply that the excitation is coherent, but very short lived.

At this juncture, it is illustrative to comment on the difference between these
results and those resulting when the equilibrated 
system (S$+$TB) is excited by a coherent source where it is assumed  
that it is the system S that interacts with the radiation. Given this coherent 
excitation (e.g. pulsed transform limited laser), 
the dynamical features change completely: the absorption 
of one photon from a coherent pulse creates a \emph{coherent} superposition of 
energy eigenstates, and hence a time evolving state (cf. Ref.~\cite{BS12b}).  
It is worth mentioning that in absence of TB, if the initial state is a pure state, then 
this superposition will be described by a \emph{pure} state.
By contrast, under the same circumstance (absence of TB), the blackbody radiation
will create a \emph{incoherent} superposition of states leading to a \emph{mixed}
state.
In the presence of TB, the stationary coherent terms created during the thermalization 
with TB will allow us to enhance  the coherent control over the populations by transferring 
the coherence of the pulse into the populations \cite{PYB13,PB13}.

As noted above, the parameters in the computation in Fig.~\ref{fig:TBBB} and 
Fig.~\ref{fig:rho11TBBB} are artificial, and designed to show some effects due to BB. 
Based on naive classical arguments, this value of $\Omega_{\mathrm{BB}}$ would be 
equivalent to the inverse of the time needed by light to travel over around 1 \AA{}.
In the context of electronic energy transfer in photosynthetic complexes, where
natural frequencies $\omega_0 = 6.63 \times 10^{12}$~Hz occur, then the coupling
constant will be  $\approx \gamma_{\mathrm{BB}} = 8\times10^{-6}\omega_0$
($\gamma_{\mathrm{BB}}^{-1} = 19$~ns). 
In this case the dynamics induced would be far less noticeable than the one shown in 
Fig.~\ref{fig:TBBB} and Fig.~\ref{fig:rho11TBBB}.
That is, since $\gamma_{\mathrm{BB}}$ is smaller, then the effect on the relaxation 
will be weaker as well. 
Indeed, the final stationary population reached after equilibration will be far closer to that 
reached during the first thermalization step.
Once again the response is to the turn-on of the field with rapid equilibration following 
mediated by the driving initial-condition-dependent-transient-term.

In summary, irradiating with blackbody radiation cannot generate
a coherent dynamical result, since the blackbody radiation is correctly
represented as a thermal bath [see Eq.~(\ref{equ:JwBB})]. 
Any time evolution observed   here  can  be  understood  as  the
ultrafast transients as the system thermalizes, mediated by 
the driving initial-condition-dependent-transient-term. This term is present
given the radiation field case, but is absent when 
excitation results from a second thermal bath coupled through the position
of the bath modes. 
\section{Discussion and Concluding Remarks}
\label{disc}


We have considered the natural process where a system, originally coupled to a thermal 
bath, is subsequently perturbed by either another thermal bath or by blackbody radiation. 
We have shown that the first step of thermalization generates off-diagonal stationary 
coherences which, \emph{in principle}, could affect the dynamics of the second 
perturbative step. 
In particular, these off-diagonal stationary coherences could affect the populations of 
states during the second step.
This process is  enhanced in the low temperature regime, and plays a fundamental 
role in explaining the origin of one photon phase control in molecular systems \cite{PYB13,PB13}. 
However, although our formal approach clearly identifies the stationary off-diagonal system 
matrix elements as participants in the subsequent time  evolution  of the populations of a 
generic open quantum system, when  these off-diagonal terms are generated \emph{in 
practice}, i.e. in natural environments or by blackbody irradiation,  their  contribution 
is negligible. 
This leads  us  to  the  conclusion  that under natural conditions,
the  off-diagonal  elements  generated by thermal baths do not
play any relevant role in the dynamics and that incoherent excitation of
an open quantum system leads to dynamics free of coherent time evolution after
an initially short transient time interval.
Incoherent dynamics do occur, however, such as heat flux between the two baths via the 
system S \cite{PB13b}.

A final note is in order. Our description of TB resembles, e.g., the role 
of a solvent. However, if the bath is part of the same macromolecule, then the spectral 
density $J_{\mathrm{TB}}(\omega)$ could be highly structured and some new features 
may be expected, e.g., increasing the stationary coherences even with weak decay rates. 
In those cases, the role of stationary coherences has to be explicitly calculated for each 
system at hand.
\begin{acknowledgements}
LAP acknowledges discussions with Prof.~Yehiam Prior and Dr. Timur V. Tscherbul with pleasure. 
This work was supported by the US Air Force Office of Scientific Research under contract
number FA9550-10-1-0260, by \textit{Comit\'e para el Desarrollo de la Investigaci\'on}
--CODI-- of Universidad de AntioquiaColombia under contract number E01651 and under the 
\textit{Estrategia de Sostenibilidad 2013-2014}, and by the \textit{Departamento Administrativo de 
Ciencia, Tecnolog\'ia e Innovaci\'on} --COLCIENCIAS-- of Colombia under the project number 
111556934912.
\end{acknowledgements}

\bibliography{praieoqsv2}

\end{document}